\documentstyle[11pt,amssymb,epsfig,psfig]{article}

\begin{document}

\title{{\bf The Casimir Effect on Background of Conformally Flat Brane--World
Geometries }}

\author{
A. A. Saharian  $^{1}$\footnote{E-mail: saharyan@www.physdep.r.am }
and  M. R. Setare $^2$ \footnote{E-mail: rezakord@yahoo.com} \\
 {$^1$ Department of Physics, Yerevan
State University, Yerevan, Armenia } \\
and
\\ {$^2$ Institute for Theoretical Physics and Mathematics, Tehran,
Iran} \\ {Department of Science, Physics Group, Kurdistan
University, Sanandeg, Iran}}

\maketitle

\begin{abstract}
The Casimir effect due to conformally coupled bulk scalar fields
on background of conformally flat brane-world geometries is
investigated. In the general case of mixed boundary conditions
formulae are derived for the vacuum expectation values of the
energy-momentum tensor and vacuum forces acting on boundaries. The
special case of the AdS bulk is considered and the application to
the Randall-Sundrum scenario is discussed. The possibility for the
radion stabilization by the vacuum forces is demonstrated.

\end{abstract}

\newpage

\section{Introduction}

The past few years witnessed a growing interest among particle
physicists and cosmologists toward models with extra space-like
dimensions. This interest was initiated by string theorists
\cite{Witt96}, who exploited a moderately large size of an
external 11th dimension in order to reconcile the Planck and
string/GUT scales. Taking this idea further, it was shown that
large extra dimensions allow for a reduction of the fundamental
higher-dimensional gravitational scale down to the TeV-scale
\cite{Arka98}. An essential ingredient of such a scenario is the
confinement of the standard model fields on field theoretical
defects, so that only gravity can access the large extra
dimensions. These models are argued to make contact with an
intricate phenomenology, with a variety of consequences for
collider searches, low-energy precision measurements, rare decays
and astroparticle physics and cosmology. However, the mechanisms,
responsible for the stabilization of extra dimensions, remain
unknown. The fact that the size of extra dimensions is large as
compared to the fundamental scale also remains unexplained. An
alternative solution to the hierarchy problem was proposed in Ref.
\cite{Rand99}. This higher dimensional scenario is based on a
non-factorizable geometry which accounts for the ratio between the
Planck scale and weak scales without the need to introduce a large
hierarchy between fundamental Planck scale and the
compactification scale. The model consists of a spacetime with a single $%
S^1/Z_2$ orbifold extra dimension. Three-branes with opposite
tension reside at the orbifold fixed points, and together with a
finely tuned negative bulk cosmological constant serve as sources
for five-dimensional gravity. The resulting spacetime metric
contains a redshift factor which depends exponentially on the
radius of the compactified dimension. In the scenario presented in
\cite{Rand99} the distance between the branes is associated with
the vacuum expectation value of a massless four dimensional scalar
field. This modulus field has zero potential and consequently the
distance is not determined by the dynamics of the model. For this
scenario to be relevant, it is necessary to find a mechanism for
generating a potential to stabilize the distance between the
branes. Classical stabilization forces due to the non-trivial
background configurations of a scalar field along an extra
dimension were first discussed by Gell-Mann and Zwiebach
\cite{Gell84}. With the revived interest in extra dimensions and
brane worlds, as modified version of this mechanism, which
exploits a classical force due to a bulk scalar field with
different interactions with the branes, received significant
attention \cite{Gold99,Gher00} (and references therein). However,
as it was shown in Refs. \cite{Kant00,Barg00}, a classical scalar
interaction is not useful for the stabilization of two positive
tension branes. Another stabilization mechanism arises due to the
Casimir force generated by the quantum fluctuations about a
constant background of a massless scalar field. For a conformally
coupled scalar this effect was initially studied in Ref.
\cite{Fabi00} in the context of M-theory, and subsequently in
Refs. \cite{Garr00,Noji00a,Noji00b,Noji00,Nayl02} for a background
Randall--Sundrum geometry (see also \cite{Seta01b} for the case of
the de Sitter bulk). Recently a variant of the brane-world model
with a compact hyperbolic manifold as a topologically non-trivial
internal space is proposed \cite{Kalop00,Trod00}. As it has been
shown in Refs. \cite{Stark01a,Stark01b} cosmology in such spaces
has interesting consequences for the evolution of the early
universe. The problem of radion stabilization in hyperbolic
brane-world scenarios is considered in \cite{Nasr02}.

In the present paper we will investigate the vacuum expectation
values of the energy--momentum tensor of the conformally coupled
scalar field on background of the conformally flat Brane-World
geometries. We will consider the general plane--symmetric
solutions of the gravitational field equations and boundary
conditions of the Robin type on the branes. The latter includes
the Dirichlet and Neumann boundary conditions as special cases.
The Casimir energy-momentum tensor for these geometries can be
generated from the corresponding flat spacetime results by using
the standard transformation formula. Previously this method has
been used in \cite{Seta01} to derive the vacuum characteristics of
the Casimir configuration on background of the static domain wall
geometry for a scalar field with Dirichlet boundary condition on
plates. For Neumann or more general mixed boundary conditions we
need to have the Casimir energy-momentum tensor for the flat
spacetime counterpart in the case of the Robin boundary conditions
with coefficients related to the metric components of the
brane-world geometry and the boundary mass terms. The Casimir
effect for the general Robin boundary conditions on background of
the Minkowski spacetime was investigated in Ref. \cite{RomSah} for
flat boundaries, and in \cite{Saha01a,Saha01b} for spherically and
cylindrically symmetric boundaries in the case of a general
conformal coupling. Here we use the results of Ref. \cite{RomSah}
to generate vacuum energy--momentum tensor for the plane symmetric
conformally flat backgrounds. The paper is organized as follows.
In the next section the vacuum expectation values of the
energy--momentum tensor and vacuum forces acting on branes are
evaluated for a general case of a conformally-flat plane symmetric
background. In section \ref{sec:AdS} the important special case of
the AdS background is considered and the possibility for the
stabilization of the distance (radion field) between the branes is
discussed. Finally, the results are re-mentioned and discussed in
section \ref{sec:conclusion}.

\section{Vacuum expectation values for the energy-momentum tensor}
\label{sec:vacemt}

In this paper we will consider a conformally coupled massless scalar field $%
\varphi (x)$ satisfying the equation
\begin{equation}
\left( \nabla _{\mu }\nabla ^{\mu }+\zeta R\right) \varphi (x)=0,\quad \zeta
=\frac{D-1}{4D}  \label{fieldeq}
\end{equation}
on background of a $D+1$--dimensional conformally flat plane--symmetric
spacetime with the metric
\begin{equation}
g_{\mu \nu }=e^{-2\sigma (z)}\eta _{\mu \nu },\quad \mu ,\nu =0,1,\ldots ,D.
\label{metric}
\end{equation}
In Eq. (\ref{fieldeq}) $\nabla _{\mu }$ is the operator of the
covariant derivative, and $R$ is the Ricci scalar for the metric
$g_{\mu \nu }$. Note that for the metric tensor from Eq.
(\ref{metric}) one has
\begin{equation}
R=De^{2\sigma }\left[ 2\sigma ^{\prime \prime }-(D-1)\sigma ^{\prime 2}%
\right] ,  \label{Riccisc}
\end{equation}
where the prime corresponds to the differentiation with respect to $z$.

We will assume that the field satisfies the mixed boundary condition
\begin{equation}
\left( a_{j}+b_{j}n^{\mu }\nabla _{\mu }\right) \varphi (x)=0,\quad
z=z_{j},\quad j=1,2  \label{boundcond}
\end{equation}
on the hypersurfaces $z=z_{1}$ and $z=z_{2}$, $z_{1}<z_{2}$,
$n^{\mu }$ is the normal to these surfaces, $n_{\mu }n^{\mu }=-1$,
and $a_j$, $b_j$ are constants. The results in the following will
depend on the ratio of these coefficients only. However, to keep
the transition to the Dirichlet and Neumann cases transparent we
will use the form (\ref{boundcond}). For the case of plane
boundaries under consideration introducing a new coordinate $y$ in
accordance with
\begin{equation}
dy=e^{-\sigma }dz  \label{ycoord}
\end{equation}
conditions (\ref{boundcond}) take the form
\begin{equation}
\left( a_{j}+(-1)^{j-1}b_{j}e^{\sigma (z_{j})}\partial _{z}\right) \varphi
(x)=\left( a_{j}+(-1)^{j-1}b_{j}\partial _{y}\right) \varphi (x)=0,\quad
y=y_{j},\quad j=1,2.  \label{boundcond1}
\end{equation}
Note that the Dirichlet and Neumann boundary conditions are
obtained from Eq. (\ref{boundcond}) as special cases corresponding
to $(a_j,b_j)=(1,0)$ and $(a_j,b_j)=(0,1)$ respectively. Our main
interest in the present paper is to investigate the vacuum
expectation values (VEV's) of the energy--momentum tensor for the field $%
\varphi (x)$ in the region $z_{1}<z<z_{2}$. The presence of
boundaries modifies the spectrum of the zero--point fluctuations
compared to the case without boundaries. This results in the shift
in the VEV's of the physical quantities, such as vacuum energy
density and stresses. This is the well known Casimir effect.

\begin{figure}[tbph]
\begin{center}
\epsfig{figure=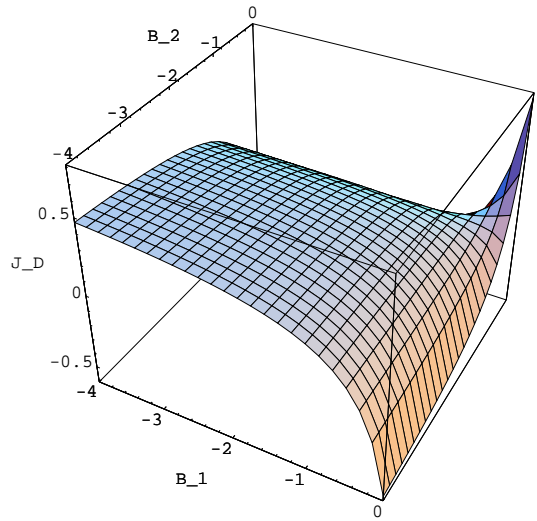,width=7cm,height=7cm}
\end{center}
\caption{ The function $J_D(B_1,B_2)$ versus $B_1$ and $B_2$ for
$D=4$. } \label{fig1ID}
\end{figure}

It can be shown that for a conformally coupled scalar by using field
equation (\ref{fieldeq}) the expression for the energy--momentum tensor can
be presented in the form
\begin{equation}
T_{\mu \nu }=\nabla _{\mu }\varphi \nabla _{\nu }\varphi -\zeta \left[ \frac{%
g_{\mu \nu }}{D-1}\nabla _{\rho }\nabla ^{\rho }+\nabla _{\mu }\nabla _{\nu
}+R_{\mu \nu }\right] \varphi ^{2},  \label{EMT1}
\end{equation}
where $R_{\mu \nu }$ is the Ricci tensor. The quantization of a scalar filed
on background of metric (2) is standard. Let $\{\varphi _{\alpha
}(x),\varphi _{\alpha }^{\ast }(x)\}$ be a complete set of orthonormalized
positive and negative frequency solutions to the field equation (\ref
{fieldeq}), obying boundary condition (\ref{boundcond}). By expanding the
field operator over these eigenfunctions, using the standard commutation
rules and the definition of the vacuum state for the vacuum expectation
values of the energy-momentum tensor one obtains
\begin{equation}
\langle 0|T_{\mu \nu }(x)|0\rangle =\sum_{\alpha }T_{\mu \nu }\{\varphi {%
_{\alpha },\varphi _{\alpha }^{\ast }\}},  \label{emtvev1}
\end{equation}
where $|0\rangle $ is the amplitude for the corresponding vacuum state, and
the bilinear form $T_{\mu \nu }\{{\varphi ,\psi \}}$ on the right is
determined by the classical energy-momentum tensor (\ref{EMT1}). In the
problem under consideration we have a conformally trivial situation:
conformally invariant field on background of the conformally flat spacetime.
Instead of evaluating Eq. (\ref{emtvev1}) directly on background of the
curved metric, the vacuum expectation values can be obtained from the
corresponding flat spacetime results for a scalar field $\bar{\varphi}$ by
using the conformal properties of the problem under consideration. Under the
conformal transformation $g_{\mu \nu }=\Omega ^{2}\eta _{\mu \nu }$ the $%
\bar{\varphi}$ field will change by the rule
\begin{equation}
\varphi (x)=\Omega ^{(1-D)/2}\bar{\varphi}(x),  \label{phicontr}
\end{equation}
where for metric (\ref{metric}) the conformal factor is given by
$\Omega =e^{-\sigma (z)}$. The boundary conditions for the field
$\bar{\varphi}(x)$ we will write in form similar to Eq.
(\ref{boundcond1})
\begin{equation}
\left( \bar{a}_{j}+(-1)^{j-1}\bar{b}_{j}\partial _{z}\right) \bar{\varphi}%
=0,\quad z=z_{j},\quad j=1,2,  \label{bounconflat}
\end{equation}
with constant Robin coefficients $\bar{a}_{j}$ and $\bar{b}_{j}$. Comparing
to the boundary conditions (\ref{boundcond}) and taking into account
transformation rule (\ref{phicontr}) we obtain the following relations
between the corresponding Robin coefficients
\begin{equation}
\bar{a}_{j}=a_{j}+(-1)^{j-1}\frac{D-1}{2}\sigma ^{\prime }(z_{j})e^{\sigma
(z_{j})}b_{j},\quad \bar{b}_{j}=b_{j}e^{\sigma (z_{j})}.  \label{coefrel}
\end{equation}
Note that as Dirichlet boundary conditions are conformally
invariant the Dirichlet scalar in the curved bulk corresponds to
the Dirichlet scalar in a flat spacetime. However, for the case of
Neumann scalar the flat spacetime counterpart is a Robin scalar
with $\bar{a}_j=(-1)^{j-1}(D-1)\sigma '(z_j)/2$ and $\bar{b}_j=1$.
The Casimir effect with boundary conditions (\ref{bounconflat}) on
two parallel plates on background of the Minkowski spacetime is
investigated in Ref. \cite{RomSah} for a scalar field with a
general conformal coupling parameter. In the case of a conformally
coupled scalar the corresponding regularized VEV's for the
energy-momentum tensor are uniform in the region between the
plates and have the form
\begin{equation}
\langle \bar{T}_{\nu }^{\mu }\left[ \eta _{\alpha \beta }\right] \rangle _{%
{\rm ren}}=-\frac{J_D(B_1,B_2)}{2^{D}\pi ^{D/2}a^{D+1}\Gamma
(D/2+1)}{\rm diag}(1,1,\ldots ,1,-D), \quad z_{1}< z< z_{2},
\label{emtvevflat}
\end{equation}
where
\begin{equation}\label{IDB1B2}
  J_D(B_1,B_2)={\rm p.v.}
\int_{0}^{\infty }\frac{t^{D}dt}{\frac{(B_{1}t-1)(B_{2}t-
1)}{(B_{1}t+1)(B_{2}t+1)}e^{2t}-1},
\end{equation}
and we use the notations
\begin{equation}
B_{j}=\frac{\bar{b}_{j}}{\bar{a}_{j}a},\quad j=1,2,\quad a=z_{2}-z_{1}.
\label{Bjcoef}
\end{equation}
For the Dirichlet and Neumann scalars $B_1=B_2=0$ and
$B_1=B_2=\infty $ respectively, and one has
\begin{equation}\label{JDDirNeu}
  J_D(0,0)=J_D(\infty ,\infty )=\frac{\Gamma (D+1)}{2^{D+1}}\zeta
  _R(D+1),
\end{equation}
with the Riemann zeta function $\zeta _R(s)$. Note that in the
regions $z< z_{1}$ and $z> z_{2}$ the Casimir densities vanish
\cite{RomSah}:
\begin{equation}
\langle \bar{T}_{\nu }^{\mu }\left[ \eta _{\alpha \beta }\right] \rangle _{%
{\rm ren}}=0,\quad z< z_{1},z> z_{2}.  \label{emtvevflat2}
\end{equation}
This can be also obtained directly from Eq. (\ref{emtvevflat})
taking the limits $z_{1}\rightarrow -\infty $ or $z_{2}\rightarrow
+\infty $. The values of the coefficients $B_{1}$ and $B_{2}$ for
which the denominator in the subintegrand of Eq.
(\ref{emtvevflat}) has zeros are specified in
\cite{RomSah}. In particular, this denominator has no zeros for $%
\{B_{1}+B_{2}\geq 1,B_{1}B_{2}\leq 0\}\cup \{B_{1,2}\leq 0\}$. In
Fig. \ref{fig1ID} the function $J_D(B_1,B_2)$ is plotted versus
$B_1$ and $B_2$ for $D=4$ and $B_{1,2}\leq 0$.

The vacuum energy-momentum tensor on curved background (\ref{metric}) is
obtained by the standard transformation law between conformally related
problems (see, for instance, \cite{Birrell}) and has the form
\begin{equation}
\langle T_{\nu }^{\mu }\left[ g_{\alpha \beta }\right] \rangle _{{\rm ren}%
}=\langle T_{\nu }^{\mu }\left[ g_{\alpha \beta }\right] \rangle _{{\rm ren}%
}^{(0)}+\langle T_{\nu }^{\mu }\left[ g_{\alpha \beta }\right] \rangle _{%
{\rm ren}}^{(b)}.  \label{emtcurved1}
\end{equation}
Here the first term on the right is the vacuum energy--momentum
tensor for the situation without boundaries (gravitational part),
and the second one is due to the presence of boundaries. As the
quantum field is conformally coupled and the background spacetime
is conformally flat the gravitational part of the energy--momentum
tensor is completely determined by the trace anomaly and is
related to the divergent part of the corresponding effective
action by the relation \cite{Birrell}
\begin{equation}
\langle T_{\nu }^{\mu }\left[ g_{\alpha \beta }\right] \rangle _{{\rm ren}%
}^{(0)}=2g^{\mu \sigma }(x)\frac{\delta }{\delta g^{\nu \sigma }(x)}W_{{\rm %
div}}[g_{\alpha \beta }].  \label{gravemt}
\end{equation}
Note that in odd spacetime dimensions the conformal anomaly is absent and
the corresponding gravitational part vanishes:
\begin{equation}
\langle T_{\nu }^{\mu }\left[ g_{\alpha \beta }\right] \rangle _{{\rm ren}%
}^{(0)}=0,\quad {\rm for\;even}\;D.  \label{gravemteven}
\end{equation}
The boundary part in Eq. (\ref{emtcurved1}) is related to the corresponding
flat spacetime counterpart (\ref{emtvevflat}),(\ref{emtvevflat2}) by the
relation \cite{Birrell}
\begin{equation}
\langle T_{\nu }^{\mu }\left[ g_{\alpha \beta }\right] \rangle _{{\rm ren}%
}^{(b)}=\frac{1}{\sqrt{|g|}}\langle \bar{T}_{\nu }^{\mu }\left[ \eta
_{\alpha \beta }\right] \rangle _{{\rm ren}}.  \label{translaw}
\end{equation}
By taking into account Eq. (\ref{emtvevflat}) from here we obtain
\begin{equation}
\langle T_{\nu }^{\mu }\left[ g_{\alpha \beta }\right] \rangle _{{\rm ren}%
}^{(b)}=-\frac{e^{(D+1)\sigma (z)}J_D(B_1,B_2)}{2^{D}\pi ^{D/2}a^{D+1}\Gamma (D/2+1)}%
{\rm diag}(1,1,\ldots ,1,-D),  \label{bpartemt}
\end{equation}
for $z_{1}< z< z_{2}$, and
\begin{equation}
\langle T_{\nu }^{\mu }\left[ g_{\alpha \beta }\right] \rangle _{{\rm ren}%
}^{(b)}=0,\;{\rm for}\;z< z_{1},z> z_{2}.  \label{bpartemt2}
\end{equation}
In Eq. (\ref{bpartemt}) the constants $B_{j}$ are related to the
Robin coefficients in boundary condition (\ref{boundcond}) by
formulae (\ref {Bjcoef}),(\ref{coefrel}) and are functions on
$z_j$. In particular, for Neumann boundary conditions
$B^{(N)}_j=2(-1)^{j-1}/[a(D-1)\sigma '(z_j)]$.

\begin{figure}[tbph]
\begin{center}
\epsfig{figure=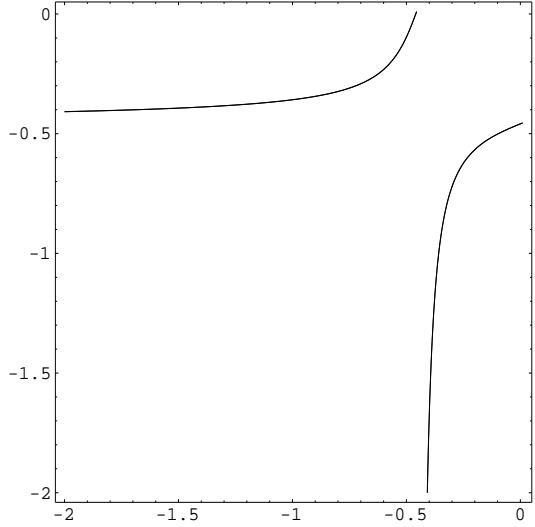,width=7cm,height=7cm}
\end{center}
\caption{ The location of zeros for the function $J_D(B_1,B_2)$ on
the quadrant $B_{1,2}\leq 0$ of the plane $(B_1,B_2)$ for $D=4$. }
\label{fig2zer}
\end{figure}

The total bulk vacuum energy per unit physical hypersurface on the
brane at $z=z_j$ is obtained by integrating over the region
between the plates
\begin{equation}\label{bulktoten}
  E_j^{(b)}=e^{D\sigma (z_j)}\int _{z_1}^{z_2}\langle T_0^0\rangle
  ^{(b)}_{{\mathrm{ren}}}e^{-(D+1)\sigma (z)}dz=-\frac{J_D(B_1,B_2)
  e^{D\sigma (z_j)}}{2^D\pi ^{D/2}\Gamma (D/2+1)a^D}.
\end{equation}
The resulting vacuum force per unit boundary area acting on the
boundary at $z=z_{j}$ is determined by the difference
\begin{equation}
\langle T_{D}^{D}\left[ g_{\alpha \beta }\right] \rangle _{{\rm ren}%
}|_{z=z_{j}+0}-\langle T_{D}^{D}\left[ g_{\alpha \beta }\right] \rangle _{%
{\rm ren}}|_{z=z_{j}-0}.  \label{vacforce0}
\end{equation}
Now we see that as gravitational part (\ref{gravemt}) is a continous
function on $z$ it does not contribute to the forces acting on the
boundaries and the vacuum force per unit surface acting on the boundary at $%
z=z_{j}$ is determined by the boundary part of the vacuum pressure, $%
p_{D}=-\langle T_{D}^{D}\left[ g_{\alpha \beta }\right] \rangle _{{\rm ren}%
}^{(b)}$, taken at the point $z=z_{j}$:
\begin{equation}
p_{Dj}(z_{1},z_{2})=-\frac{e^{(D+1)\sigma
(z_{j})}J_D(B_1,B_2)}{2^{D-1}\pi ^{D/2}a^{D+1}\Gamma (D/2)}.
\label{vacforce}
\end{equation}
This corresponds to the attractive/repulsive force between the plates if $%
p_{Dj}</>0$. The equilibrium points for the plates correspond to
the zero values of Eq. (\ref{vacforce}): $p_{Dj}=0$. These points
are zeros of the function $J_D(B_1,B_2)$ defined by Eq.
(\ref{IDB1B2}) and are the same for both plates. Note that at
these points the VEV's of the bulk energy-momentum tensor given by
Eq. (\ref{bpartemt}) and the total bulk energy also vanish. The
location of these zeros on the quadrant $B_{1,2}\leq 0$ of the
plane $(B_1,B_2)$ for $D=4$ is plotted in Fig. \ref{fig2zer}. The
function $J_D(B_1,B_2)$ is positive in the region between the
curves (including the point $(0,0)$) and is negative outside of
this region. When $B_1=0$ the zero for the function $J_4(B_1,B_2)$
is obtained for $B_2\approx -0.459$. In the limit $B_1\to \infty $
for the zero one has $B_2\approx -0.451$.

\section{Casimir densities and vacuum forces on AdS background}
\label{sec:AdS}

\begin{figure}[tbph]
\begin{center}
\begin{tabular}{ccc}
\epsfig{figure=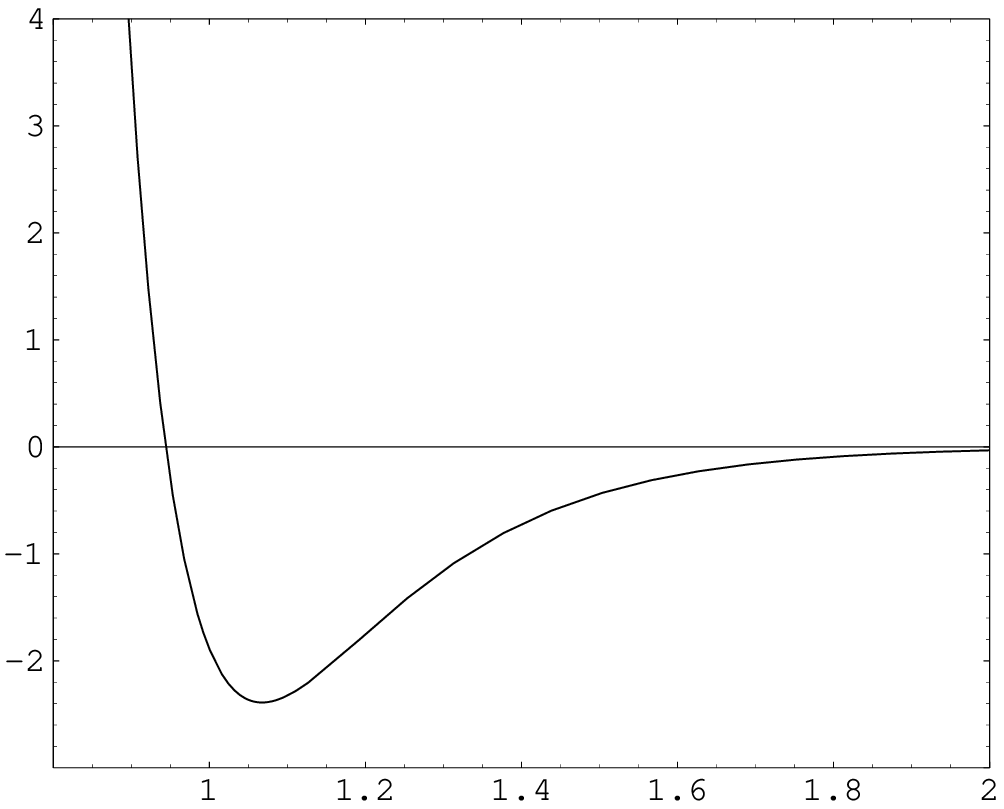,width=6cm,height=6cm} & \hspace*{0.5cm} & %
\epsfig{figure=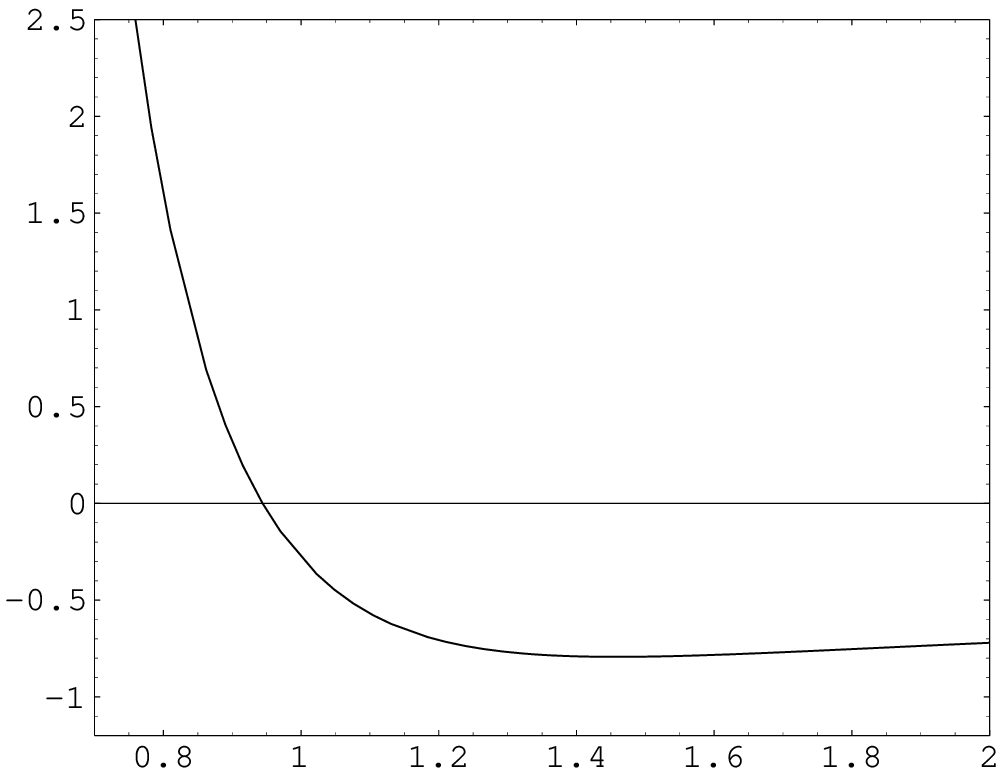,width=6cm,height=6cm}
\end{tabular}
\end{center}
\caption{ The vacuum forces acting per unit surfaces of the branes
on the AdS background, $10^5p_{D1}/k_D^{D+1}$ (left panel) and
$10^2p_{D2}/k_D^{D+1}$ (right panel) for $D=4$ versus $k_D\Delta
y$. } \label{fig3ads}
\end{figure}

As an application of the general formulae from the previous
section here we consider the important special case of the
AdS$_{D+1}$ bulk for which
\begin{equation}\label{sigads}
  \sigma =\ln (k_Dz)=k_Dy,
\end{equation}
with $1/k_D$ being the AdS curvature radius. Now the expressions
for the coefficients $B_j$, $j=1,2$ take the form
\begin{equation}
B_{j}=\frac{b_{j}k_Dz_j}{(z_2-z_1)\left[a_{j}+(-1)^{j-1}(D-1)k_{D}b_{j}/2\right]
}. \label{Bjads}
\end{equation}
Note that the ratio $z_2/z_1$ is related to the proper distance
between the branes $\Delta y$ by the formula
\begin{equation}\label{distads}
  z_2/z_1=e^{k_D\Delta y},\quad \Delta y=y_2-y_1.
\end{equation}
For the boundary induced part of the vacuum energy-momentum tensor
one has
\begin{equation}
\langle T_{\nu }^{\mu }\left[ g_{\alpha \beta }\right] \rangle
_{{\rm ren} }^{(b)}=- \left( \frac{k_Dz}{z_2-z_1}\right) ^{D+1}
\frac{J_D(B_1,B_2)}{2^{D} \pi ^{D/2}\Gamma (D/2+1)} {\rm
diag}(1,1,\ldots ,1,-D), \label{bpartemtads}
\end{equation}
where $B_j$ are functions on the distance between the branes. The
bulk vacuum energy per unit hypersurface on the brane $z=z_j$ is
obtained from (\ref{bulktoten}):
\begin{equation}\label{totenRS}
  E_j^{(b)}=-\left( \frac{k_Dz_j}{z_2-z_1}\right) ^{D}
\frac{J_D(B_1,B_2)}{2^{D} \pi ^{D/2}\Gamma (D/2+1)}.
\end{equation}
The vacuum forces per unit surface are functions on the proper
distance $\Delta y$ and on the ratio $a_j/b_j$ of the Robin
coefficients:
\begin{equation}
p_{Dj}=- \left( \frac{k_Dz_j}{z_2-z_1}\right) ^{D+1}
\frac{J_D(B_1,B_2)}{2^{D-1}\pi ^{D/2}\Gamma (D/2)},\quad j=1,2.
\label{vacforceads}
\end{equation}
In Fig. \ref{fig3ads} we have plotted the vacuum forces acting per
unit surface of the branes as functions of $k_D \Delta y$ for the
values of parameters $a_1/k_Db_1=-3$, $a_2/k_Db_2=-1$ and $D=4$.
As we see from these graphics there is an equilibrium point at
$\Delta y_0\approx 0.95/k_D$, where the vacuum forces vanish.
These forces are repulsive for $\Delta y<\Delta y_0$ ($p_{Dj}>0$)
and are attractive for $\Delta y>\Delta y_0$ ($p_{Dj}<0$). As a
result the equilibrium point is stable. Hence, we have an example
for the stabilization of the distance between the plates due to
the vacuum forces.

Now we turn to the brane--world model introduced by Randall and
Sundrum \cite{Rand99} and based on the AdS geometry with one extra
dimension. The fifth dimension $y$ is compactified on an orbifold,
$S^1/Z_2$ of length $\Delta y$, with $-\Delta y\leq y\leq \Delta
y$. The orbifold fixed points at $y=0$ and $y=\Delta y$ are the
locations of two 3-branes. For the conformal factor in this model
one has $\sigma =k_D|y|$. The boundary conditions for the
corresponding conformally coupled bulk scalars have the form
(\ref{boundcond1}) with Robin coefficients $a_j/b_j=-c_jk_D$,
where the constants $c_j$ are the coefficients in the boundary
mass term \cite{Gher00}:
\begin{equation}\label{boundmass}
  m_{\varphi }^{(b)2}=2k_D\left[ c_1\delta (y)+c_2\delta (y-\Delta
  y)\right] .
\end{equation}
Note that here we consider the general case when the boundary
masses are different for different branes. Supersymmetry requires
$c_2=-c_1$. The boundary induced part of the VEV's for the
energy-momentum tensor on the Randall-Sundrum brane-world
background are obtained from Eq. (\ref{bpartemtads}) with
additional factor 1/2. This factor is related to the fact that now
in the normalization condition for the eigenfunctions the
integration goes over the region $(-\Delta y,\Delta y)$, instead
of $(0,\Delta y )$. The expressions for the total energy and
vacuum forces remain the same. The coefficients $B_j$ in the
expression for $J_D(B_1,B_2)$ are given by the formula (below we
will keep $D$ general)
\begin{equation}\label{BjRS}
  B_j=-\frac{e^{(j-1)k_D\Delta y}}{e^{k_D\Delta y}-1}\frac{1}{c_j
  +(-1)^{j}(D-1)/2}.
\end{equation}
In the Randall-Sundrum model, the hierarchy problem is solved if
$k_D\Delta y\approx 37$. For these values of distances between the
branes one has $|B_1|\ll |B_2|$ under the assumption $c_1\sim
c_2$.

Expression (\ref{totenRS}) takes into account the Casimir energy
of the bulk. In general, the total vacuum energy receives
additional contribution coming from the vacuum energy located on
branes. In the case of general mixed boundary conditions the
decomposition of the vacuum energy into surface and volume parts
is presented in Ref. \cite{RomSah} for the Minkowski background.
The corresponding results for a conformally coupled scalar on the
AdS bulk is obtained by a way similar to that described above. For
the total vacuum energy per unit hypersurface on the brane $z=z_j$
this yields
\begin{equation}\label{totenRSglob}
  E_j^{(b){\mathrm{tot}}}=-\left( \frac{k_Dz_j}{z_2-z_1}\right) ^{D}
\frac{2^{-D-1} \pi ^{-D/2}}{\Gamma (D/2+1)}{\mathrm{p.v.}}\int
_0^\infty dt \, t^D\frac{d}{dt}\ln \left[
1-\frac{(B_1t+1)(B_2t+1)}{(B_1t-1)(B_2t-1)}e^{-2t}\right] .
\end{equation}
In addition, the vacuum energy per unit hypersurface on the brane
$z=z_j$ can contain terms in the form $k_D^D\sum _{l=1}^{2} \alpha
_l(z_j/z_l)^D $ with constants $\alpha _1$ and $\alpha _2 $ and
corresponding to the single brane contributions when the second
brane is absent. Adding these terms to the vacuum energy
corresponds to finite renormalization of the tension on both
branes (see \cite{Garr00,Noji00} for more detailed discussion).

\section{Conclusion}
\label{sec:conclusion}

In the present paper we have investigated the Casimir effect for a
conformally coupled scalar field confined in the region between
two parallel branes on background of the conformally-flat plane
symmetric spacetimes. The general case of the mixed boundary
conditions is considered. The vacuum expectation values of the
energy-momentum tensor are derived from the corresponding flat
spacetime results by using the conformal properties of the
problem. The purely gravitational part arises due to the trace
anomaly and is zero for odd spacetime dimensions. In the region
between the branes the boundary induced part for the vacuum
energy-momentum tensor is given by formula (\ref{bpartemt}), and
the corresponding vacuum forces acting per unit surface of the
brane have the form Eq. (\ref{vacforce}). These forces vanish at
the zeros of the function $J_D(B_1,B_2)$. In the case $B_{1,2}\leq
0 $ the subintegrand in Eq. (\ref{vacforce}) has no real poles and
the location of these zeros for $D=4$ is plotted in Fig.
\ref{fig2zer}. Further we consider a special case of the AdS bulk
with the brane induced vacuum energy-momentum tensor given by Eq.
(\ref{bpartemt}). On a specific example we demonstrate that there
are stable equilibrium points, where the vacuum forces vanish and
the radion field is stabilized. An application to the
Randall-Sundrum brane-world model is discussed. In this model the
coefficients in the Robin boundary conditions on branes are
related to the boundary mass terms for the scalar field under
consideration. In the present paper we have considered the
geometry of flat branes. The corresponding results for spherical
branes can be obtained by applying the same method of conformal
transformation to the corresponding flat spacetime vacuum
energy-momentum tensor given in \cite{Saha01a} (for the total
Casimir energy in the cases of spherical one and two brane
configurations see Refs. \cite{Noji00b,Nayl02}).

\section*{Acknowledgement }

We acknowledge support from the Research Project of the Kurdistan
University. The work of AAS was supported in part by the Armenian
Ministry of Education and Science (Grant No. 0887).


\begin{thebibliography}{99}
\bibitem{Witt96}  E. Witten, Nucl. Phys. {\bf B471}, 135 (1996); P.
Horava and E. Witten, Nucl. Phys. {\bf B460}, 506 (1996); T. Banks
and M. Dine, Nucl. Phys. {\bf B479}, 173 (1996).

\bibitem{Arka98}  N. Arkani-Hamed, S. Dimopoulos, and G. Dvali, Phys. Lett.
{\bf B429}, 263 (1998); Phys. Rev. {\bf D59}, 086004 (1999); I.
Antoniadis, N. Arkani-Hamed, S. Dimopoulos, and G. Dvali, Phys.
Lett. {\bf B436}, 257 (1998).

\bibitem{Rand99}  L. Randall and R. Sundrum, Phys. Rev. Lett. {\bf 83}, 3370
(1999).

\bibitem{Gell84}  M. Gell-Mann and B. Zwiebach, Phys. Lett. {\bf B141}, 333
(1984); Nucl. Phys. {\bf B260}, 569 (1985).

\bibitem{Gold99}  W. D. Goldberger and M. B. Wise, Phys. Rev. Lett. {\bf 83},
4922 (1999); Phys. Rev. {\bf D60}, 107505 (1999); C. Csaki, J.
Erlich, T. Hollowood, and Y. Shirman, Nucl. Phys. {\bf B581}, 309
(2000); K. Maeda and D. Wands, Phys. Rev. {\bf D62}, 124009
(2000); C. Barcelo and M. Visser, Phys. Rev. {\bf D63}, 024004
(2001).

\bibitem{Gher00} T. Gherghetta and A. Pomarol, Nucl. Phys. {\bf
B586}, 141 (2000).

\bibitem{Kant00}  P. Kanti, K. A. Olive, and M. Pospelov, Phys. Lett. {\bf
B481},386 (2000).

\bibitem{Barg00} V. Barger, T. Han, T. Li, J. D. Lykken, and D. Marfatia,
Phys. Lett. {\bf B488}, 97 (2000).

\bibitem{Fabi00}  M. Fabinger and P. Horava, Nucl. Phys. {\bf B580}, 243
(2000).

\bibitem{Garr00}  J. Garriga, O. Pujolas, and T. Tanaka, Nucl.
Phys. {\bf B605}, 192 (2001).

\bibitem{Noji00a} S. Nojiri, S. Odintsov, and S. Zerbini, Phys.
Rev. {\bf D62}, 064006 (2000).

\bibitem{Noji00b} S. Nojiri, S. Odintsov, and S. Zerbini, Class.
Quantum Grav. {\bf 17}, 4855 (2000).

\bibitem{Noji00}  S. Nojiri and S. Odintsov, Phys. Lett. {\bf B484}, 119
(2000); S. Nojiri, O. Obregon, and S. Odintsov, Phys. Rev. {\bf
D62}, 104003 (2000); D. J. Toms, Phys. Lett. {\bf B484}, 149
(2000); W. Goldberger and I. Rothstein, Phys. Lett. {\bf B491},
339 (2000); S. Nojiri and S. Odintsov, JHEP {\bf 0007}, 049
(2000); I. Brevik, K. A. Milton, S. Nojiri, and S. D. Odintsov,
Nucl. Phys. {\bf B599}, 305 (2001); A. Flachi and D. J. Toms,
Nucl. Phys. {\bf B599}, 305 (2001); R. Hofmann, P. Kanti, and M.
Pospelov, Phys. Rev. {\bf D63}, 124020 (2001).

\bibitem{Nayl02} W. Naylor and M. Sasaki, Phys. Lett. {\bf B542},
289 (2002).

\bibitem{Seta01b} M. R. Setare and R. Mansouri, Class. Quantum
Grav. {\bf 18}, 2695 (2001).

\bibitem{Kalop00} N. Kaloper, J. March-Russel, G. D. Starkman, and
M. Trodden, Phys. Rev. Lett. {\bf 85}, 928 (2000).

\bibitem{Trod00} M. Trodden, Diluting gravity with compact hyperboloids,
hep-th/0010032.

\bibitem{Stark01a} G. D. Starkman, D. Stojkovic, and M. Trodden, Phys. Rev.
Lett. {\bf 87}, 231303 (2001).

\bibitem{Stark01b} G. D. Starkman, D. Stojkovic, and M. Trodden, Phys. Rev.
{\bf D63}, 103511 (2001).

\bibitem{Nasr02} S. Nasri, P. J. Silva, G. D. Starkman, and M. Trodden, Radion
stabilization in compact hyperbolic extra dimensions, hep-th/0201063.

\bibitem{Seta01}  M. R. Setare and A. A. Saharian, Int. J. Mod. Phys. {\bf A16},
1463 (2001).

\bibitem{RomSah}  A. Romeo and A. A. Saharian, J. Phys. A:
Math. Gen., {\bf \ 35}, 1297 (2002).

\bibitem{Saha01a}  A. A. Saharian, Phys. Rev. {\bf D63}, 125007 (2001).

\bibitem{Saha01b}  A. Romeo and A. A. Saharian, Phys. Rev. {\bf D63}, 105019
(2001).

\bibitem{Birrell} N. D. Birrel and P. C. W. Davies, {\it Quantum Fields in
Curved Space} (Cambridge: Cambridge University Press, 1982).

\end{thebibliography}
\end{document}